\documentclass[9pt,twocolumn,twoside]{opticajnl}
\journal{opticajournal} % use for journal or Optica Open submissions

\setboolean{shortarticle}{true}
\usepackage{lineno}
\usepackage{siunitx}
\newcommand{\SIadj}[2]{\SI[number-unit-product={\text{-}}]{#1}{#2}}
\usepackage{jabbrv}

\title{Polarization rotation through differential transmission in refractive CMB telescopes identified using a hybrid physical optics method}

\author[1,*]{Xiaodong Ren}
\author[1]{Rustam Balafendiev}
\author[1,2]{Jon E. Gudmundsson}

\affil[1]{Science school, University of Iceland, Sæmundargata 2, 102 Reykjavík, Iceland}
\affil[2]{The Oskar Klein Centre, Department of Physics, Stockholm University, SE-106 91 Stockholm, Sweden}
\affil[*]{xren@hi.is}

\begin{abstract}
We identify a polarization rotation systematic in the far field beams of refractive cosmic microwave background (CMB) telescopes caused by differential transmission in anti-reflection (AR) coatings of optical elements. This systematic was identified following the development of a hybrid physical optics method that incorporates full-wave electromagnetic simulations of AR coatings to model the full polarization response of refractive systems. Applying this method to a two-lens CMB telescope with non-ideal AR coating, we show that polarization-dependent transmission can produce a rotation of the far-field polarization angle that varies across the focal plane with a typical amplitude of $0.05-\SI{0.5}{\degree}$. If ignored in analysis, this effect can produce temperature to polarization leakage and Stokes $Q/U$ mixing. 
\end{abstract}

\setboolean{displaycopyright}{false} % Do not include copyright or licensing information in submission.

\begin{document}

\maketitle

\section{Introduction}
 Refractive optical systems are critical for modern cosmic microwave background (CMB) experiments \cite{Takahashi2010, cmbs4-technology, SOScience_2019} due to their on-axis configuration and ability to house optical elements and detector arrays in a compact cryogenic environment. This combination results in highly symmetric beams, low cross-polarization, a large diffraction-limited field of view (FoV), and reduced thermal radiation loading from the warm environment. Small-aperture instruments such as the BICEP series \cite{Aikin2010}, the Simons Observatory Small Aperture Telescopes \cite{SOSAT_Telescope_paper}, and the balloon-borne SPIDER experiment \cite{Rahlin2014} therefore employ fully refractive designs to measure large-scale CMB polarization and search for a primordial $B$-mode signal \cite{BicepKeck2021}. For experiments that require both high angular resolution and high optical throughput, refractive optics remain essential but appear as cold back-optics combined with a dual-reflector fore-optics system, as in ACT \cite{Fowler2007}, the South Pole Telescope \cite{Padin2008}, FYST \cite{parshley2018ccat, parshley2018optical, huber2024ccat}, and the Simons Observatory Large Aperture Telescope \cite{Gudmundsson:21, Zhu_2021}. In these hybrid architectures, the refractive optics play the role of shaping the detector beams and further correcting aberrations introduced by the reflective elements to support a wide FoV.

 The science goals of current and future CMB telescopes place stringent requirements on the beam and polarization response characterization. Driven by searches for primordial $B$-modes and cosmic birefringence \cite{Minami2020}, the polarization angle of upcoming CMB experiments needs to be known to better than $0.1^\circ$. Beam and polarization imperfections such as cross-polarization, position-dependent polarization rotation, and beam asymmetries can generate temperature-to-polarization leakage and Stokes $Q/U$ mixing that may bias cosmological parameter estimation \cite{shimon2008cmb,Fraisse2013}. Therefore, accurate calibration of the polarization angle is critical and relies on a combination of electromagnetic modeling, optical metrology, and point source observations.  In this paper, we describe a state-of-the-art physical optics simulation algorithm and use it to investigate a thus far unidentified polarization rotation systematic in refractive microwave telescopes. The systematic effect is produced by polarization-dependent transmission in optical elements.

 %\textcolor{red}{\sout{In this paper, we focus on the modeling aspect of this effort and develop a state-of-the-art physical optics simulation algorithm for refractive optics, enabling precise prediction of polarization rotation effects.}}\textcolor{blue}{In this paper, we investigate a beam-level polarization rotation systematic in on-axis refractive CMB telescopes, focusing on polarization angle rotation induced by polarization-dependent transmission in the optical elements.}

 To model the polarization response of a refractive optical system to sky signals, one can either use fast geometrical optics methods or more complete wave-based approaches such as physical optics (PO) analysis \cite{collin1969antenna} and physical theory of diffraction (PTD) technique \cite{ufimtsev2014fundamentals}. Geometrical optics analysis implements polarization ray tracing algorithm, e.g. the \textit{Poldsp} algorithm used in CODE V and the polarization ray tracing and polarization pupil map analyses in ZEMAX, to follow the evolution of the polarization state along individual rays. They can capture polarization rotation and attenuation induced by tilted optical elements, birefringent materials, and coating surfaces of the optical elements, and thus provide useful first-order estimates of how the optical layout affects polarization \cite{Murphy:24}. However, ray tracing neglects diffraction and wave-interference effects and as a result this method cannot predict the detailed co- and cross-polar beam patterns and diffraction-induced polarization rotation. For quantitative studies of polarization systematics, PO and PTD simulations are required to model the full beam profile and study the polarization response of the optics.

 Refractive optical systems almost always require anti-reflection (AR) coatings \cite{Datta:13,Golec:22} on their lens and filter surfaces to suppress reflection losses. Accurate and efficient modeling of the optical performance of such systems while incorporating the effects of realistic AR coatings, however, remains challenging. Commercial PO software packages, most notably TICRA Tools (aka GRASP), are widely used for the electromagnetic analysis of large reflector and quasi-optical systems, and have been successfully applied to the design and verification of many millimeter and sub-millimeter reflective instruments \cite{tauber2010planck}. When applied to refractive elements, however, the standard PO analysis or simple lens-PO functionality provided by GRASP treats the refractive interfaces as bare dielectric surfaces and simply multiplies the field at the interfaces by Fresnel coefficients, without accounting for the full, frequency- and angle-dependent response of multi-layer AR coatings. On the other hand, full-wave electromagnetic solvers can model AR coatings with high fidelity, but their computational cost becomes prohibitive for large-aperture, multi-lens CMB systems. As a result, there is currently no practical tool capable of performing efficient PO beam propagation while simultaneously including realistic AR-coating effects. Motivated by this, we propose a hybrid PO approach, in which AR coatings of the optical elements are expressed by Jones matrices, obtained from full-wave electromagnetic (EM) simulations, and incorporated at the refractive surfaces within the PO propagation chain. 
 
 The paper is organized as follows. In Section 2, we describe the details of the proposed hybrid-PO simulation approach for refractive optics and define the polarization angle conventions used in this paper. In Section 3, we introduce a wide-field of view two-lens CMB refractor telescope and its corresponding lens AR coating design. The telescope serves as a representative model and as a testbed for validating the proposed hybrid-PO approach. Section 4 first assesses the accuracy of the hybrid-PO method by comparing its predictions with state-of-the-art full-wave EM simulations, and then uses this method to study the polarization systematics of the two-lens optics. This section identifies a polarization rotation effect caused by differential transmission in refractive elements which, until now, has not been described in the literature. The main conclusions are summarized in Section 5.  

%-------------------------------------------------------------------
\section{Optical Modeling Methods}
\label{section:method}
\subsection{Hybrid Physical Optics Analysis for Refractive Optics} \label{section:po}
 In a real CMB instrument, the faint signals originating from the sky are either collected by large reflector antennas (typically a primary and secondary reflector) and delivered to cold refractive optics, or directly collected by a purely refractive optical system before finally focusing the incoming light onto polarization-sensitive detectors in the focal plane. To predict the far field beam patterns of these telescopes, however, it is more convenient to model the time-reversed process: the transmission pattern, in which the field is launched from the detector feedhorn and propagated backward through the optics. This approach is justified by reciprocity theory \cite{collin1969antenna}. PO analysis is the standard technique used to simulate the beam patterns for microwave telescopes and its accuracy has been validated for large reflective systems, for example in pre-launch beam modeling of the \textit{Planck} telescope \cite{tauber2010planck}. The application of PO to dielectric lens analysis has also been demonstrated in the literature \cite{A_PO, mom_po_go, DPO_BORMOM}. In this work, the authors even employ a double PO (DPO) method which combines the main PO propagation of the transmitted field with first order effects from internal reflections to improve the overall accuracy of the analysis, with the results showing good agreement with full-wave (method of moments) solutions. 

 % inside the lens are explicitly taken into account and a second PO propagation is applied to the reflected field
 
 Unfortunately, in both conventional PO and DPO methods the dielectric lens surfaces are still modeled as bare dielectric interfaces, which does not allow the effects of realistic anti-reflection (AR) coatings to be included. As a result, one cannot study how uncertainties in AR-coating thicknesses or refractive indices propagate into the optical beams, nor quantify the corresponding polarization systematics induced by this effect. To address this limitation, we adopt a hybrid PO method in which AR coating layers are described by polarization-dependent Jones matrices derived from full-wave EM simulations that act as field operators at each refractive interface within the PO propagation.
 
 The AR coating is treated using a local plane-wave approximation: at each point on the curved lens surface, the coating response is assumed to be identical to that of the same layered structure implemented on an infinite planar surface and illuminated by a plane wave at the corresponding incidence angle. The EM performance of the AR coating can be efficiently modeled by full-wave simulations, for example unit-cell simulation with Floquet periodic boundary conditions implemented in \textit{CST Studio Suite}. The responses of the AR coating layer for \textit{s}- and \textit{p}-polarized waves are represented by two \text{$2\times2$} Jones matrices, one for transmission $\mathbf{J}_{t}$ and one for reflection $\mathbf{J}_{r}$. In the case of an isotropic ARC, both matrices are diagonal:
 \begin{align}
     \mathbf{J}_{t}(\nu,\theta) &= 
        \begin{bmatrix}
         t_{ss}(\nu,\theta) & 0 \\ %T_{TE,TM}
         0     & t_{pp}(\nu,\theta) %T_{TM,TE} 
        \end{bmatrix}, \label{eq:jones_matrix_a}\\
     \mathbf{J}_{r}(\nu,\theta) &= 
        \begin{bmatrix}
         r_{ss}(\nu,\theta) & 0 \\
         0      & r_{pp}(\nu,\theta)
        \end{bmatrix}, \label{eq:jones_matrix_b}
    %\label{eq:jones_matrix}
 \end{align}
 where $t_{ss}$, $t_{pp}$, $r_{ss}$ and $r_{pp}$ are the complex transmission and reflection coefficients for $s$ and $p$ polarization as functions of frequency $\nu$ and incident angle $\theta$. These coefficients are complex-valued to account for both the amplitude and the phase delay introduced by the AR coating. In practice, they are precomputed on a grid in $(\nu,\theta)$ and interpolated during the PO propagation. 
 
 Other thin transmissive elements in the optical chain, such as filters and the vacuum window, which are often realized as thin dielectric slabs, can be treated in the same way by inserting their corresponding Jones matrices at the appropriate planes in the PO propagation. We note that, in principle, time-varying polarization elements such as a rotating half-wave plate can also be represented within the same hybrid-PO framework by a time-dependent Jones matrix $\mathbf{J}(\nu, \theta,t)$ at the plate location, where the time dependence enters through the plate orientation angle. This possibility is mentioned here only to indicate the generality of the method and is not implemented in the present work.

 \begin{figure}[h]
  \centerline{\includegraphics[width=1\linewidth]{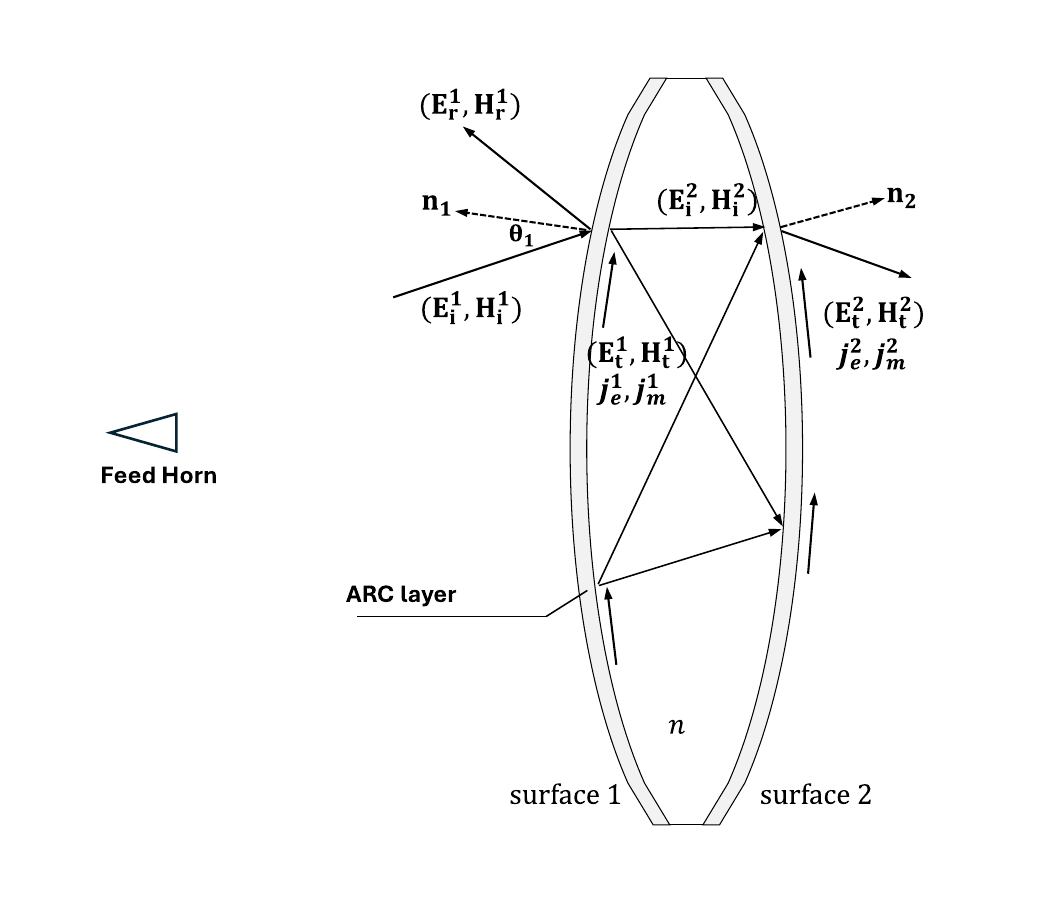}}
  \caption{Schematic showing critical parameters used in a hybrid PO calculation for a lens with anti-reflection coating on both sides.}
  \label{fig:arc_po_flow}
 \end{figure}

 Figure \ref{fig:arc_po_flow} illustrates a simple example of the PO analysis procedure for a single-lens system and how the AR coating is incorporated into the calculation. We assume that the left surface of the lens is illuminated by a source, for example a Gaussian beam launched from the detector feed. The incident fields, $\mathbf{E}^{1}_{i}$ and $\mathbf{H}^{1}_{i}$, at each point on this surface can be projected onto the local $s$- and $p$-polarization basis, which is defined by the surface normal vector and the Poynting vector of the incident field. The resulting $s$- and $p$-polarization components are then multiplied by the pre-computed transmission Jones matrix $\mathbf{J}_{t}(\nu,\theta)$ of the AR coating to obtain the transmitted fields $\mathbf{E}^{1}_{t}$ and $\mathbf{H}^{1}_{t}$. These transmitted fields are converted into equivalent electric and magnetic surface currents, $\textbf{\textit{j}}^{1}_{e}$ and $\textbf{\textit{j}}^{1}_{m}$. The currents are then used in the standard PO integral to compute the field on the right surface of the lens $\mathbf{E}^{2}_{i}$ and $\mathbf{H}^{2}_{i}$. Repeating the same procedure, the equivalent surface currents $\textbf{\textit{j}}^{2}_{e}$ and $\textbf{\textit{j}}^{2}_{m}$ can be found to compute the radiation field of the lens in the far field or on the surface of next optical element. The detailed numerical implementation of this approach, including efficient surface sampling method for the PO integrals and GPU acceleration, will be presented in a forthcoming paper on the open-source software implementation. 

\subsection{Effective Polarization Angle of Far-field Beams}
 The hybrid-PO approach enables detailed prediction of the radiation beam of refractive optics and thus provide the basis for investigating the polarization properties of refractive CMB telescopes. In order to quantify the polarization rotation of an optical system, it is necessary to establish a consistent method for defining the polarization direction of the far-field beam, since multiple definitions exist in the literature.

 The far-field beam electric field, denoted by $\mathbf{E}_{\text{far}}$, can always be completely expressed as a linear combination of two mutually orthogonal polarization components on the sky, both of which are perpendicular to the radiation direction. We choose the two polarization basis vectors according to Ludwig's third definition \cite{ludwig_1973}. The corresponding two unit vectors are denoted co-polar ($\hat{\mathbf{e}}_{\mathrm{co}}$) and cross-polar ($\hat{\mathbf{e}}_{\mathrm{cx}}$). The far-field beam can then be written as:
    \begin{equation}
        \mathbf{E}_{\text{far}} = E_{\mathrm{co}}\,\hat{\mathbf{e}}_{\mathrm{co}} + E_{\mathrm{cx}}\,\hat{\mathbf{e}}_{\mathrm{cx}},
    \end{equation}
 with
    \begin{align}
        \hat{\mathbf{e}}_{\mathrm{co}} &= \hat{\theta}\ \cos \phi - \hat{\phi}\ \sin \phi, \\
        \hat{\mathbf{e}}_{\mathrm{cx}} &= \hat{\theta}\ \sin \phi + \hat{\phi}\ \cos \phi,
    \end{align}
 where $\hat{\theta}$ and $\hat{\phi}$ are the spherical unit vectors of the adopted coordinate system. In this reference frame, the polarization direction of the far-field beam is defined as the angle by which the original sky reference frame is rotated to maximize some kind of ratio between the rotated co-polar and cross-polar components. There are three approaches introduced in \cite{franco2003systematic} to compute the beam polarization angle, $\varphi_{\mathrm{beam}}$. We adopt the 3rd method, which determines the angle by maximizing the ratio between the integrated co-polar and cross-polar fields over the main beam given by
 \begin{equation}
     \varphi_\mathrm{beam} = \arg\max_{\varphi} \frac{\int_{\Omega_0}|E^{\varphi}_\mathrm{co}|^2 d\Omega}{\int_{\Omega_0}|E^{\varphi}_\mathrm{cx}|^2 d\Omega},
 \end{equation}
 where $\Omega_0$ denotes the main beam region, defined here as the set of directions where the field power exceeds \SI{-20}{\deci\bel} relative to the beam peak. The fields $E^{\varphi}_\mathrm{co}$ and $E^{\varphi}_\mathrm{cx}$ are the co- and cross-polar components in the sky basis rotated by $\varphi$ with respect to the original reference frame.

%-------------------------------------------------------------------
\section{A Two-lens CMB Telescope}
\label{sec:twolens}
\begin{figure}[t!]
   \centerline{ \includegraphics[width=3.5in]{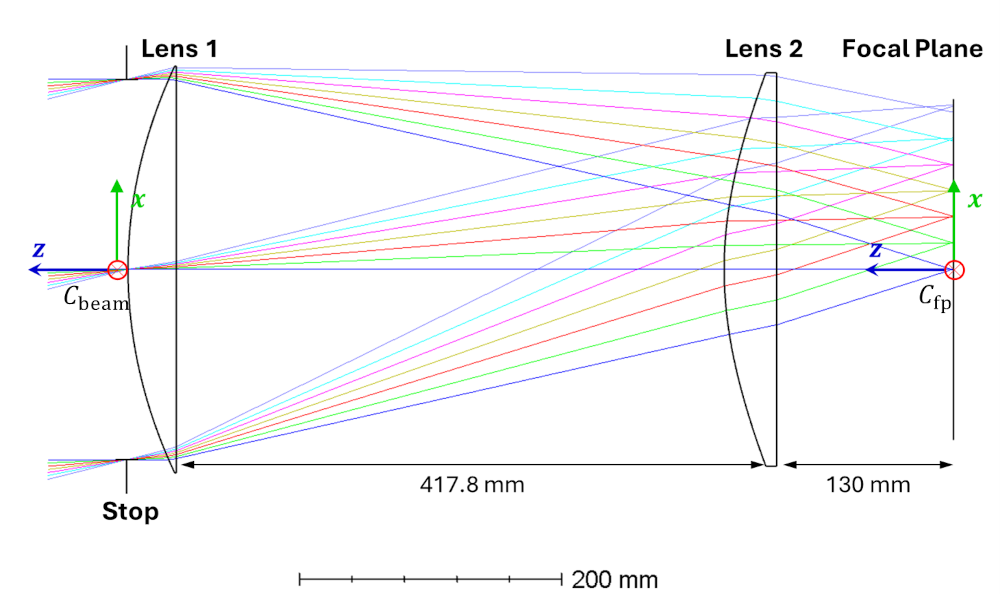}}
    \caption{Optical schematic of the two-lenses telescope \cite{gudmundsson2020geometrical}. The two lenses have a diameter of 300 mm and are constructed of HDPE. $C_{\mathrm{fp}}$ is the coordinate system on the focal plane. $C_{\mathrm{beam}}$ indicates the far-field coordinate system.}
    \label{Fig:Two_lens_optics}
\end{figure}

In this section we describe a two-lens CMB optical system used as a testbed for the optical modeling framework developed in Section~\ref{section:method}. The optical design follows the configurations developed in \cite{gudmundsson2020geometrical}, where it was presented as part of a comparative study of refractive optics for CMB observations, contrasting the performance of silicon and plastic lenses. In this work, we adopt the plastic design, which consists of two plano-convex lens. Both lenses have a diameter of 300 mm and are made of high-density polyethylene (HDPE). The refractive index of HDPE is assumed to be $n = 1.52$ at all designed frequencies. Figure \ref{Fig:Two_lens_optics} illustrates the optical layout. The optics was designed and optimized to provide a theoretical diffraction-limited field of view of $\pm14^{\circ}$ for the 90- and \SIadj{150}{\giga\hertz} bands, which corresponds to a circular area of approximately $\pm \SI{120}{\milli\metre}$ on the focal plane. The details of the optical design and optimization procedure are described in paper \cite{gudmundsson2020geometrical}.
%\subsection{Anti-Reflection Coating}

\begin{figure}[t!]
  \centering
  \centerline{ \includegraphics[width=3.5in]{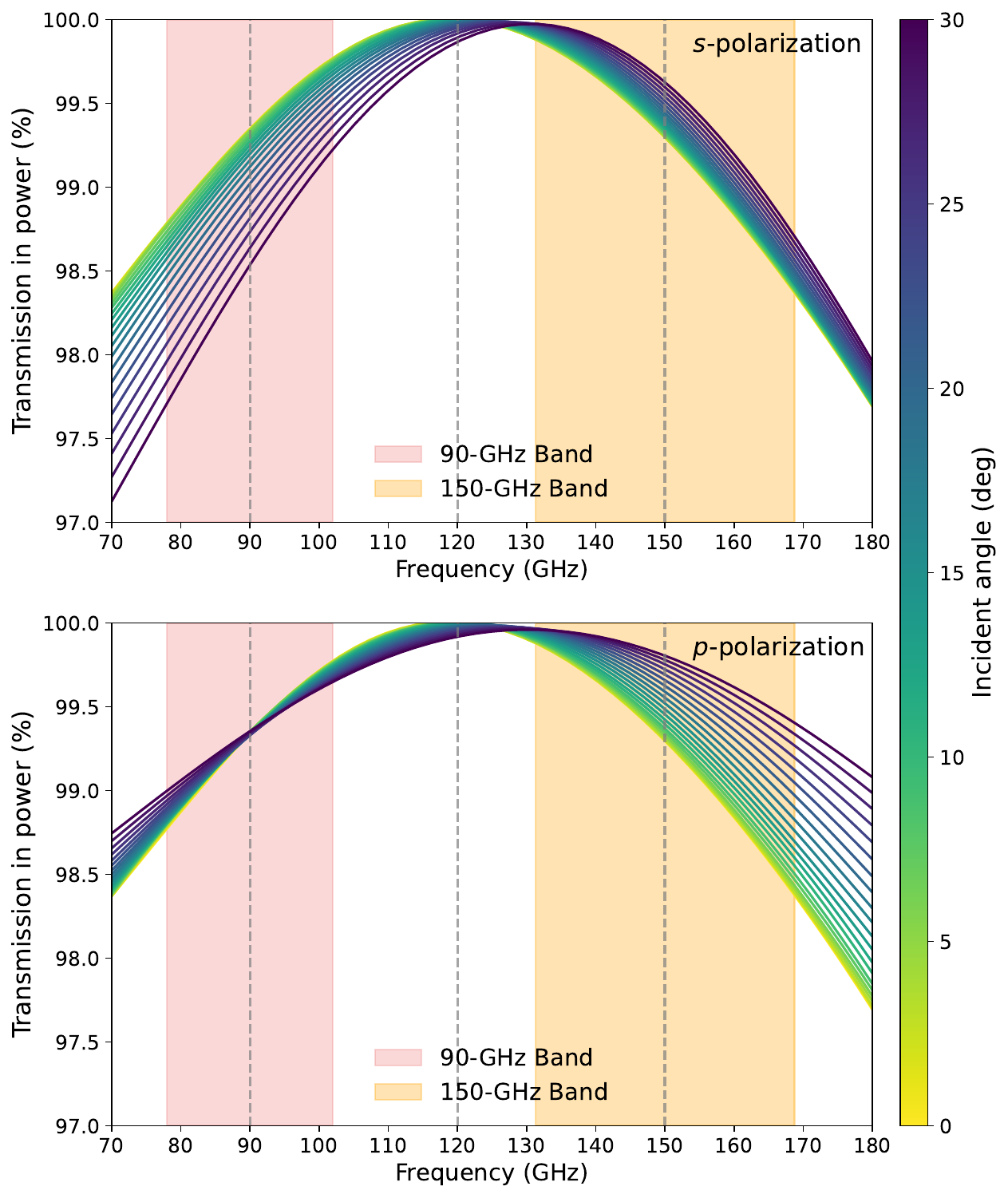}}
  \caption{ Transmission of the quarter-wave AR coating on HDPE for $s-$ (top) and $p$-polarization (bottom). Each panel shows the transmission as a function of frequency for incidence angles from $0^{\circ}$ (normal incidence) to $30^{\circ}$. The curve color indicates the incidence angle according to the colorbar. The coating has an effective refractive index of $n_\mathrm{AR}=1.23$ and a physical thickness of \SI{0.51}{\milli\metre}.}
  \label{fig:arc_perf}
\end{figure}

Both lenses are coated with a single-layer AR coating to improve transmission efficiency and suppress polarization-dependent reflections, which would otherwise cause $\sim4\%$ reflective loss per surface. The ARC is modeled as a homogeneous dielectric layer with an effective refractive index of $n_\mathrm{AR}=\sqrt{n}$ and a thickness of \SI{0.51}{\milli\metre}, corresponding to a quarter-wave design at 120 GHz. Materials with this index, such as porous PTFE sheets (e.g. Zitex), are commonly used in millimeter-wave optics. The transmission performance of this coating is summarized in Figure~\ref{fig:arc_perf}, showing its response from 70 to \SI{180}{\giga\hertz} for both $s$- and $p$-polarization waves, for incidence angles from normal incidence up to $30^{\circ}$. The resulting reflection and transmission coefficients are incorporated into the hybrid-PO analysis. In principle, it is possible to use a two-layer AR coating with appropriately chosen refractive indices and thicknesses to achieve low reflection in both \SI{90}{\giga\hertz} and \SI{150}{\giga\hertz} bands \cite{eiben2024multilayer}. However, the detailed design and optimization of such dual-band coatings are beyond the scope of this paper. Here we adopt the simpler single-layer design to focus on verifying the hybrid-PO framework and its application to polarization-rotation analysis.
%-------------------------------------------------------------------
\section{Simulations}
\label{sec:sims}
In what follows, we apply the hybrid-PO method to the two-lens telescope to compute its far-field beam patterns and investigate its polarization properties. To simplify the simulations, we model the beam of detector feed as an ideal Gaussian beam, which means the cross-polarization of the detector is zero, so that any cross-polarization and polarization-rotation effects in the far field are generated solely by the optics. We adopt a Gaussian beam with a full width at half maximum (FWHM) of \SI{18.5}{\degree}, which produces an illumination taper of \SI{-10}{\decibel} at the aperture plane of the optics and suppresses the side-lobe to below \SI{-20}{\decibel} relative to the peak of the main beam. In the following and before the polarization study, we first examine the accuracy of the developed hybrid-PO approach.

\subsection{Accuracy of the hybrid PO method}
We evaluate the accuracy of the hybrid-PO method by comparing the simulated far-field beams of the two-lens optics with the results obtained from TICRA GRASP using the MoM analysis. The beams are computed at \SI{120}{\giga\hertz} for two representative feed positions: one at the center of the focal plane and another positioned at the edge of the focal plane, offset by \SI{120}{\milli\meter} from the optical axis. For both methods, the far-field beams are computed over an $\SI{8}{\degree}\times\SI{8}{\degree}$ patch on the sky, covering the main beam and approximately seven side-lobes.
\begin{figure}[t!]
  \centering
  \includegraphics[width=1\linewidth]{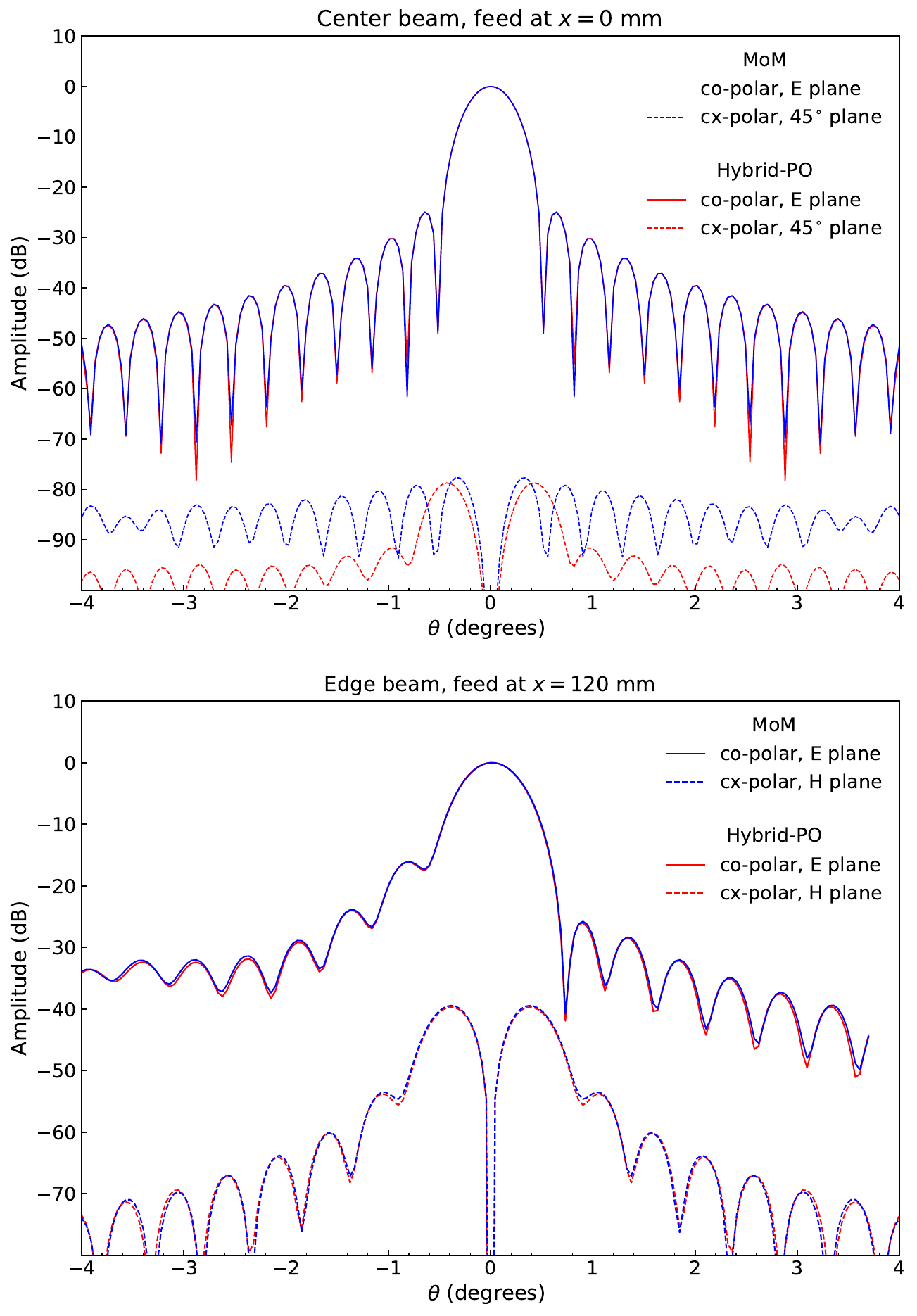}
  \caption{Far-field beam comparison of the two-lens optics at \SI{120}{\giga\hertz} obtained using hybrid-PO and MoM. Top panel shows the co- and cross-polarization beams for a feed at the focal-plane center. Bottom panel shows the corresponding co- and cx-polarization beams for a feed located at $x = \SI{120}{mm}$ in the focal planes.}
 \label{Fig:ARC_POvsMoM}
\end{figure}

Figure \ref{Fig:ARC_POvsMoM} shows the co- and cross-polar beam patterns for the center and edge feeds. The agreement in the co-polar beams is very good: the differences are below about \SI{-65}{\deci\bel} relative to the beam peak for the center beam and \SI{-40}{\deci\bel} for the edge beam. For the on-axis configuration, the cross-polarization is slightly underestimated by the hybrid-PO analysis. However, since the cross-polarization level of the center beam is nearly \SI{-80}{\deci\bel} below the co-polar peak, this discrepancy is negligible and does not affect the polarization systematics discussed in the rest of the paper. For the off-axis configuration, where the cross-polarization rises to \SI{-40}{\deci\bel}, the hybrid-PO and MoM results remain in good agreement.
    
 \subsection{Polarization offset in far-field beam}
 \label{sec:poloffset}
    This section studies the polarization properties of the optics for its far-field radiation beams. To quantify the relation between the feed orientation and the corresponding far-field beam polarization plane, reference coordinate systems are first established to define both the feed orientation and the resulting beam polarization. The polarization state and position of the Gaussian feed are specified in the focal-plane coordinate system, $C_\mathrm{fp}$, as illustrated in Figure \ref{Fig:Two_lens_optics}, where the origin is located at the center of the focal plane. The reference feed polarization direction is initially aligned with the $x$-axis, corresponding to a polarization angle of \SI{0}{\degree}. The feed rotation angle, $\varphi_\mathrm{feed}$, is defined as a rotation of the feed about its symmetry axis, which is aligned with the $z$-axis of the $C_\mathrm{fp}$ coordinate system, following the right-hand rule. In turn, the resulting polarization rotation in far-field beam is expressed by $\varphi_\mathrm{beam} - \varphi^{i}_\mathrm{beam}$, where $\varphi_\mathrm{beam}$ and $\varphi^{i}_\mathrm{beam}$ denote the far-field polarization angles corresponding to the rotated and the initial feed orientations, respectively. These angles are computed using the method described in Section~2.2 within the coordinate system $C_\mathrm{beam}$, also shown in Figure~\ref{Fig:Two_lens_optics}.
    
    In the following simulations we analyze the case in which the feed moves along the $x$-axis in the focal plane. Owing to the rotational symmetry of the optics, results for any other point at the same radius can be obtained by rotating the coordinate system by the corresponding azimuthal angle so that the testing point is mapped onto the x-axis. The polarization direction is then recovered by applying the same rotation to the far-field beam. 
    
    Ideally, rotating the feed by $\varphi_\mathrm{feed}$ from its initial orientation, the polarization angle of the far-field beam should be also rotated by the same angle, so that $\varphi_\mathrm{beam} - \varphi^{i}_\mathrm{beam} = \varphi_\mathrm{feed}$. In this case there is no relative polarization offset. However, for a non-ideal optical system, such as a system with mechanical deformations, misalignment between optical components, or polarization-dependent transmission and reflection from optical elements, optical symmetries can be broken which results in the far-field polarization angle deviating from its ideal orientation. The difference of the polarization is defined as the beam polarization offset: 
    \begin{equation}
        \Delta\varphi_\mathrm{beam} = (\varphi_\mathrm{beam}-\varphi^{i}_\mathrm{beam}) - \varphi_\mathrm{feed}. %\nonumber
                             %= \varphi_\mathrm{beam} - \varphi_\mathrm{feed}.
    \end{equation}
    To study this effect, we have carried out a series of intensive numerical simulations to probe how feed position and imperfection of AR coating impact the polarization performance of the optics covering the 90- and \SIadj{150}{\giga\hertz} bands. 
    %The polarization behavior of the far-field beam is analyzed by rotating the feed polarization around its symmetry axis in \SI{15}{\degree} increments from \SI{0}{\degree} to \SI{180}{\degree}, thereby enabling a systematic characterization of the resulting polarization offsets.
    %We particularly emphasize that the developed framework also provides a general approach for analyzing other type refractive optics operating at millimeter and sub-millimeter bands.
    %--------------------------------------------------------------------
    
\begin{figure}[t!]
        \centering
        \includegraphics[width=1\linewidth]{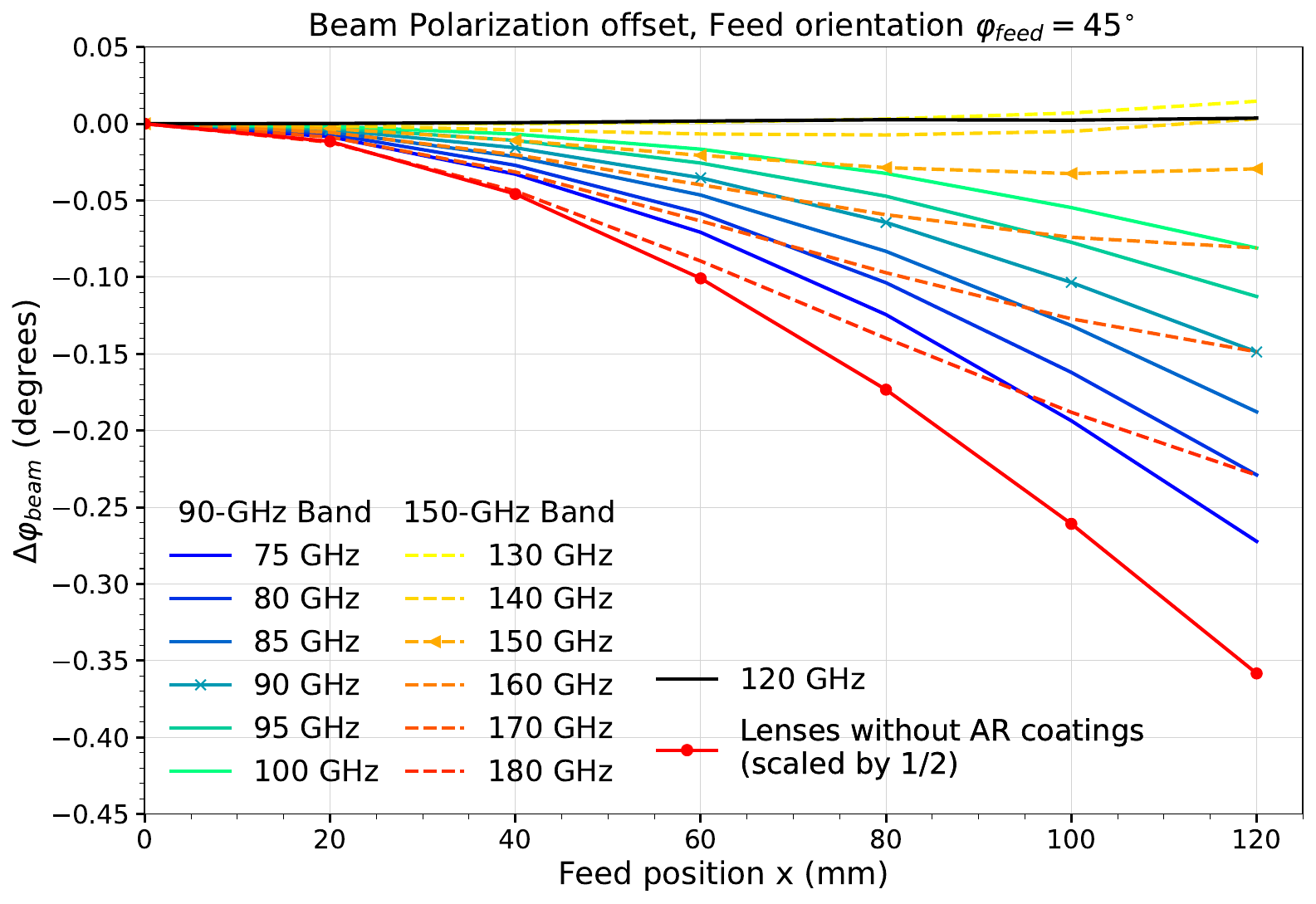}
        \caption{Far-field beam polarization offsets of the HDPE two-lens optical system as a function of feed position for the 75-100 GHz and 130-180 GHz bands, and at 120 GHz. In the optical setup, the feed is rotated by $45^{\circ}$ about its symmetry axis to evaluate the polarization response.}
        \label{fig:45_AR_90_150GHz}
    \end{figure}

    When the feed polarization direction is oriented along the $x$-axis ($\varphi_\mathrm{feed}=0^{\circ}$) or $y$-axis ($\varphi_\mathrm{feed}=90^{\circ}$), the resulting beam polarization remains aligned with the feed orientation regardless of the feed's displacement from the optical axis. This is expected because the optical system is mirror symmetric with respect to the $xz-$plane. Therefore, no rotation of the polarization basis occur for these two feed orientations. To investigate polarization offsets, the intermediate feed orientation, $\varphi_\mathrm{feed}=45^{\circ}$, is selected. Figure~\ref{fig:45_AR_90_150GHz} shows the beam polarization offset as a function of feed position along the $x$-axis in the 90- and \SIadj{150}{\giga\hertz} bands, as well as at the single frequency of \SI{120}{\giga\hertz}. At \SI{120}{\giga\hertz}, where the AR coating is optimized, the polarization offset remains below \SI{0.01}{\degree} across the full focal-plane range.
    In contrast, in the 90- and \SIadj{150}{\giga\hertz} bands, the amplitude of the polarization offset gradually increases with feed displacement from the optical axis and reaches approximately $0.3^{\circ}$ near the band edges. The systematic effect is strongest at the edge of the bands, at 75 and \SI{180}{\giga\hertz} in this particular case.
    %Since the AR coating used in the optics is optimized at \SI{120}{\giga\hertz} and low cross-polarization of the on-axis optics, the polarization offset at the frequency is minimal, less then \SI{0.01}{\degree} over the entire focal plane. 
    
    For comparison, simulations were also performed for the same optical configuration without AR coatings on the lens surfaces. The resulting polarization offsets are shown in Figure~\ref{fig:45_AR_90_150GHz}. In this case we find that the offsets are nearly independent of frequency and the maximum value reaches about $0.7^{\circ}$ for the feed located at \SI{120}{\milli\meter}. 
    %\textbf{\color{red}{45 degree angle changes.}}
    %\textcolor{blue}{show intersection view from the back of the optics and show the polarization orientations.}
    
    These simulations indicate that the offsets mainly originate from the imbalance in the transmission amplitude of the \textit{s}- and \textit{p}-polarized fields at the lens surfaces. A well-designed AR coating can significantly reduce this imbalance and thereby suppress the resulting polarization offset. For example, implementing a multi-layer AR coating and jointly optimizing the layer indices and thicknesses to achieve high transmission while minimizing the transmission difference between the two orthogonal polarizations waves at the center frequencies of the 90 and \SI{150}{\giga\hertz} would further mitigate these polarization offsets. We note that multi-layer AR coatings are for example implemented in the silicon lenses of the Simons Observatory \cite{Golec:22}. 

\subsection{Spurious optical polarization}
    Based on the above polarization offset study, when this telescope is used for polarization measurement, the resulting polarization offsets, together with the presence of cross-polarization and beam distortions, can produce spurious polarized signals that are not present in the sky. These signals are purely instrumental and must be either suppressed or accurately characterized to enable high-precision polarization measurements.
    
    \begin{figure}[t!]
       \centerline{ \includegraphics[width=3.5in]{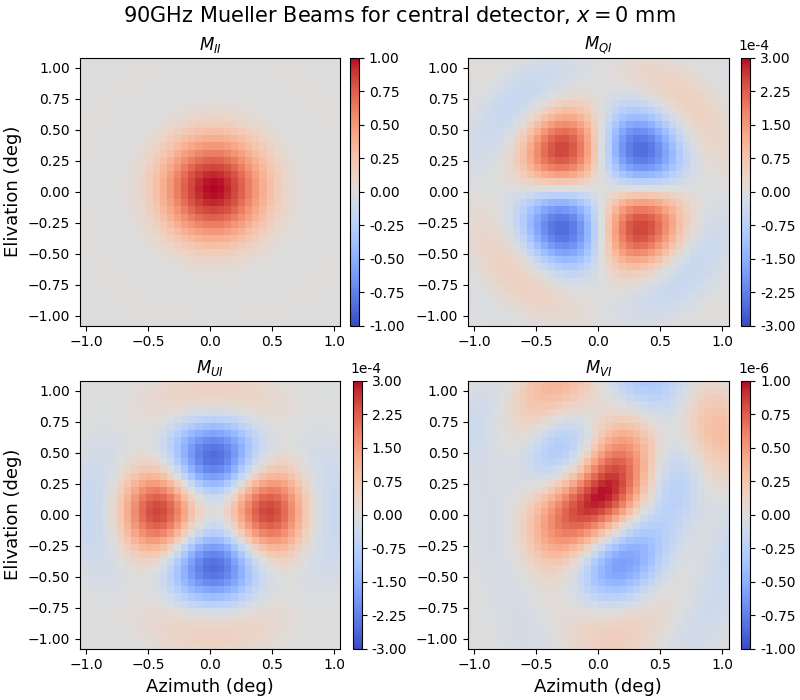}}
        \caption{ Four normalized Mueller Beams, $M_{II}$, $M_{QI}$, $M_{UI}$ and $M_{VI}$ at \SI{90}{\giga\hertz} for the optics with detector located at the center of the focal plane center, corresponding to $x=\SI{0}{mm}$. All beams are normalized to the peak of the peak of $M_{II}$.}
        \label{Fig:Stokes_beam_center}
    \end{figure}
    \begin{figure}[h]
       \centerline{ \includegraphics[width=3.5in]{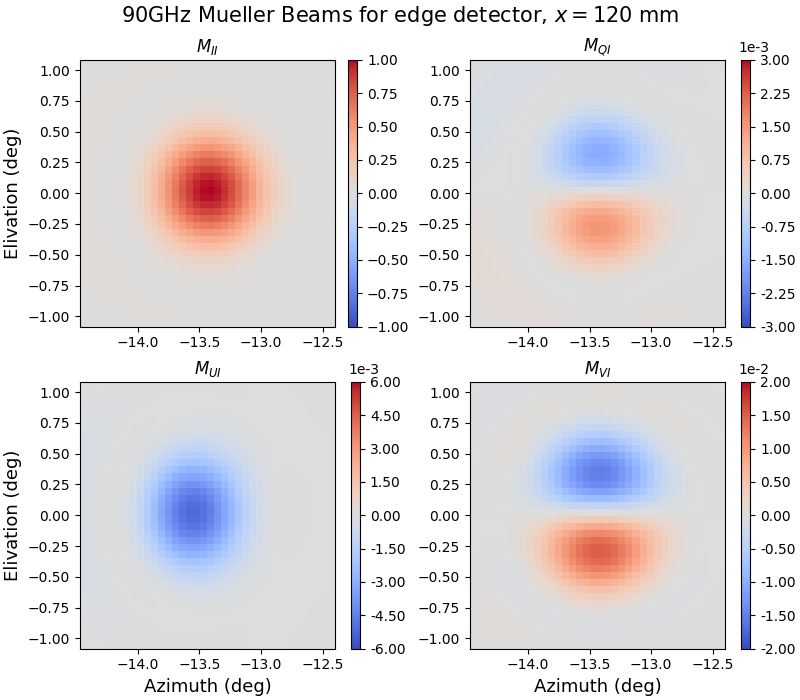}}
        \caption{Four normalized Mueller Beams at \SI{90}{\giga\hertz} with detector at the focal plane edge, corresponding to $x=\SI{120}{mm}$. All beams are normalized to the peak of the peak of $M_{II}$.}
        \label{Fig:Stokes_beam_edge}
    \end{figure}

    To quantify the impact of these optical imperfections on the polarization measurement, we describe the sky signal by the Stokes vector $S_\mathrm{sky} = (I_\mathrm{sky},Q_\mathrm{sky},U_\mathrm{sky},V_\mathrm{sky})^{\mathrm{T}}$ \cite{born2013principles}, and simulate the response of the optics to the signal. Specifically, we compute the far-field beam of the optics for a pair of linearly-polarized detectors whose polarization axes are oriented at $\pm\SI{45}{\degree}$ and use those calculations to construct the corresponding Mueller beam matrix, following the approach described by \cite{o2007systematic} and 
    \cite{jones2007instrumental}. This Mueller matrix characterizes how the optical system transforms the incoming sky Stokes parameters into those measured by the detector pair, that is
    \begin{equation}
            S_\mathrm{det} = 
                \begin{pmatrix}
                M_{II} & M_{IQ} & M_{IU} &M_{IV}\\
                M_{QI} & M_{QQ} & M_{QU} &M_{QV}\\
                M_{UI} & M_{UQ} & M_{UU} &M_{UV}\\
                M_{VI} & M_{VQ} & M_{VU} &M_{VV}\\
                \end{pmatrix}
                \cdot S_\mathrm{sky}.
    \end{equation}
    Each element of the matrix is a function of the sky angular direction. For an isotropic, unpolarized diffuse sky signal ($Q_\mathrm{sky}=U_\mathrm{sky}=V_\mathrm{sky}=0$), the response simplifies considerably: $M_{II}$ represents the total intensity beam, while $M_{QI}$ and $M_{UI}$ quantify the temperature-to-polarization leakage induced by optical imperfections. The $M_{VI}$ term describes the coupling from total intensity to circular polarization, which is typically negligible since most CMB instruments do not measure Stokes~$V$.
 
    Figure~\ref{Fig:Stokes_beam_center} shows the four Mueller beam components at \SI{90}{\giga\hertz} for the detector located at center of focal plane. All Mueller beams are normalized to the peak of the intensity map, $(M_{II,\mathrm{peak}})$. Owing to the rotational symmetry of this configuration, $M_{II}$ is nearly circular, and the temperature-to-polarization leakage terms $M_{QI}$ and $M_{UI}$ exhibit the expected clover-like pattern with peak amplitude below $2.5\times10^{-4}$ (\SI{-36}{\decibel}) relative to $M_{II,\mathrm{peak}}$ and a beam-averaged leakage consistent with zero, where the beam-average leakage, $\epsilon_{T\rightarrow P}$, is defined as the ratio of the solid-angle-integrated Mueller beam components,
    \begin{equation}
        \epsilon_{T\rightarrow P} = \frac{\int_{\Omega_0} M_{QI} d\Omega}{\int_{\Omega_0} M_{II} d\Omega}.
    \end{equation}
    
    We also consider the off-axis configuration with feed at $x = \SI{120}{\milli\metre}$ in the focal plane. As shown in Section~\ref{sec:sims}\ref{sec:poloffset}, a polarization offset of about \SI{0.15}{\degree} is present for this configuration originating from the imperfect AR coating at \SI{90}{\giga\hertz}. With this analysis, we now see that the $M_{UI}$ beam exhibits a non-zero monopole component (see Figure \ref{Fig:Stokes_beam_edge}). The resulting beam-averaged temperature-to-polarization leakage is about \SI{0.5}{\percent}.

%-------------------------------------------------------------------
\section{Conclusions}
In this work, we have described a novel hybrid physical-optics method for modeling polarization systematics in refractive CMB optical systems. Compared to existing physical optics analysis techniques, this approach accurately incorporates the effects of AR coatings on the surfaces of optical elements in the propagation chain. The AR coatings are characterized using full-wave electromagnetic simulations, converted into frequency- and angle-dependent Jones matrices that are then sampled at the corresponding optical surfaces so that their polarization-dependent transmission and phase are propagated consistently through the system. In contrast the full-wave electromagnetic analysis for large optical systems, the hybrid-PO method requires substantially less computational effort with comparable accuracy. The accuracy of the method has been verified by implementing the most accurate MoM analysis for a two-lens HDPE telescope.

We apply this new framework to investigate the polarization properties of a typical two-lens optical system for detectors located at different positions in the focal plane. We identify a new polarization systematic caused by differential transmission and find that the polarization offset grows with the feed displacement from the optical axis. This effect is caused by difference in transmission of the $s$- and $p$-polarization waves as a result of non-ideal AR coatings. Optimizing the AR coating design at the target frequency can substantially reduce the polarization offset. However, for broadband observations such as CMB experiments, the polarization offsets generated near the band edges cannot be neglected. We calculate the corresponding Mueller matrix beams to characterize the far field response corresponding to this polarization systematic. For each pixel, at \SI{90}{\giga\hertz}, \SI{0.15}{\degree} polarization offset results in temperature-to-polarization leakage at the $\sim\SI{0.5}{\percent}$ level, which is non-negligible for CMB experiments focusing on $B$-mode polarization. Knowledge of this effect should influence detector and focal plane architecture designs for future microwave telescopes and be incorporated in cosmological analysis.

%-------------------------------------------------------------------
\section{acknowledgment}
Funded in part by the European Union (ERC, CMBeam, 101040169). JEG gratefully acknowledges support from the University of Iceland Research Fund and the Icelandic Research Fund (Grant number: 2410656-051). We thank Miranda Eiben for useful discussions.
%\end{acknowledgements}

% Bibliography
\bibliography{bibliography}

% Full bibliography added automatically for Optics Letters submissions; the following line will simply be ignored if submitting to other journals.
% Note that this extra page will not count against page length
%\bibliographyfullrefs{sample0}

%Manual citation list
%\begin{thebibliography}{1}
%\bibitem{Zhang:14}
%Y.~Zhang, S.~Qiao, L.~Sun, Q.~W. Shi, W.~Huang, %L.~Li, and Z.~Yang,
 % \enquote{Photoinduced active terahertz metamaterials with nanostructured
  %vanadium dioxide film deposited by sol-gel method,} Opt. Express \textbf{22},
  %11070--11078 (2014).
%\end{thebibliography}

% Please include bios and photos of all authors for aop articles
\end{document}